\documentclass[useAMS,usenatbib]{mn2e}

\usepackage{graphicx}
\usepackage{float}
\usepackage{amssymb} 
\usepackage{bm}
\def\mpc{h^{-1} {\rm{Mpc}}}

\def\lsim{\mathrel{\hbox{\rlap{\hbox{\lower4pt\hbox{$\sim$}}}\hbox{$<$}}}}
\def\gsim{\mathrel{\hbox{\rlap{\hbox{\lower4pt\hbox{$\sim$}}}\hbox{$>$}}}}

\newcommand{\aap}    {A\&A}

\newcommand{\apj}    {ApJ}

\newcommand{\prd} {Phys. Rev. D}
\newcommand{\mnras}  {MNRAS}
\newcommand{\jcap}  {JCAP}

\title
[Improving the precision matrix for precision cosmology]
{Improving the precision matrix for precision cosmology}

\author[Paz \& S\'anchez]{
\parbox[t]{\textwidth}
{ 
  Dante J. Paz$^{1,2}$ \thanks{E-mail: dpaz@oac.unc.edu.ar}
  and Ariel G. S\'anchez$^{3}$
}
\vspace*{6pt}\\
$^1$ Instituto de Astronom\'\i a Te\'orica y Experimental, 
     UNC-CONICET, C\'ordoba, Argentina. \\
$^2$ Observatorio Astron\'omico de C\'ordoba, Universidad Nacional de C\'ordoba, Argentina. \\
$^3$ Max-Planck-Institut f\"ur extraterrestrische Physik, Postfach 1312, Giessenbachstr., 85741 Garching, Germany.\\ 
}

\begin{document}

\date{Submitted to MNRAS}

\maketitle

\begin{abstract} 
	The estimation of cosmological constraints from observations of the
	large scale structure of the Universe, such as the power spectrum or
	the correlation function, requires the knowledge of the inverse of the
	associated covariance matrix, namely the precision matrix,
	$\mathbf{\Psi}$.  In most analyses, $\mathbf{\Psi}$  is estimated from
	a limited set of mock catalogues.  Depending on how many mocks are
	used, this estimation has an associated error which must be propagated
	into the final cosmological constraints. For future surveys such as
	Euclid and DESI, the control of this additional uncertainty requires a
	prohibitively large number of mock catalogues.  In this work we test a
	novel technique for the estimation of the precision matrix, the
	covariance tapering method, in the context of baryon acoustic
	oscillation measurements.  Even though this technique was originally
	devised as a way to speed up maximum likelihood estimations, our
	results show that it also reduces the impact of noisy precision matrix
	estimates on the derived confidence intervals, without introducing
	biases on the target parameters.  The application of this technique can
	help future surveys to reach their true constraining power using a
	significantly smaller number of mock catalogues.  
\end{abstract} 


\pagebreak
\section{Introduction}
\label{S_intro}

The small statistical uncertainties associated with current cosmological
observations allow for precise cosmological constraints to be derived
\citep[e.g.][]{2014Anderson, Planck_2015}.  Future stage IV experiments such as
Euclid \citep{euclid_paper} and the Dark Energy Spectroscopic Instrument
\citep[DESI][]{desipaper} will push the attainable level of precision even
further, providing a strong test of the standard $\Lambda$CDM cosmological
model.

As statistical uncertainties are reduced, the control of potential systematic
errors becomes essential to derive robust cosmological constraints. Besides the
correct treatment of the observations and accurate models of the data,
precision cosmology requires a thorough control of the assumptions made when
establishing the link between theory and observations. For example, most
analyses of clustering statistics assume a Gaussian likelihood function.  This
assumption must be carefully revised as they might introduce systematic biases
on the obtained confidence levels \citep{2015Kalus}.

Even for the Gaussian case, the evaluation of the likelihood function requires
the knowledge of the precision matrix, ${\bf \Psi}$, that is, the inverse of
the covariance matrix of the measurements.  In most analyses of clustering
measurements, the precision matrix is estimated from an ensemble of mock
catalogues reproducing the selection function of each survey
\citep[e.g.][]{2013Manera, 2014Manera}.  However, all estimates of ${\bf \Psi}$
based on a finite number of mock catalogues are affected by noise.  A rigorous
statistical analysis requires the propagation of these uncertainties into the
final cosmological parameter constraints.

Recent studies have provided a clear description of the dependence of the noise
in the estimated precision matrix on the number of mock catalogues used
\citep*{2013Taylor}, its propagation to the derived parameter uncertainties
\citep{2013Dodelson,2014Taylor} and the correct way to include this additional
uncertainty in the obtained cosmological constraints \citep{2014Percival}.  The
results from these studies show that, depending on the number of bins of a
given measurement, a large number of mock catalogues might be necessary in
order to keep this additional source of uncertainty under control.  For future
large-volume surveys such as Euclid or DESI, this requirement might be
infeasible, even if these are based on approximated methods such as
\textsc{pinocchio} \citep{2002Monaco}, \textsc{cola}
\citep{2013Tassev,2015Koda}, \textsc{patchy} \citep{2014Kitaura} or
\textsc{ezmocks} \citep{2015Chuang} instead of full $N$-body simulations.

In this paper we test the implementation of the Covariance Tapering (CT) method
\citep*{kaufman_covariance_2008} as a tool to minimize the impact of the noise
in precision matrix estimates derived from a finite set of mock catalogues.
Although this technique was originally designed as a way to speed-up the
calculation of maximum likelihood estimates, we show that this method can also
be used to reduce the noise in the estimates of the precision matrix.  The
covariance tapering approach can help to obtain parameter constraints that are
close to ideal (i.e.  those derived when the true covariance matrix is known)
even when the precision matrix is estimated from a manageable number of
realizations.  Our results show that CT can significantly reduce the number of
mock catalogues required for the analysis of future surveys, allowing these
data to reach their full constraining power.

The structure of the paper is as follows. In Sec.~\ref{bao_cov} we summarize
the results of previous works regarding the impact of the noise in precision
matrix estimates on cosmological constraints.  The CT technique is described in
Sec.~\ref{taper}.  In Sec.~\ref{ct_practice} we apply CT to the same test case
studied by \citet{2014Percival}, that of normally-distributed independent
measurements of zero mean. We then extend this analysis to a case with non-zero
intrinsic covariances.  Section~\ref{app_bao} presents a study of the
applicability of CT to the measurement of the baryon acoustic oscillations
(BAO) signal using Monte Carlo realizations of the of the large-scale two-point
correlation function.  Finally, in Sect.~\ref{conclusions} we present our main
conclusions.

\section{Impact of precision matrix errors on cosmological constraints}
\label{bao_cov}

In most cosmological analyses the information from observations is compressed
into a measurement ${\bf D}$, such as the power spectrum or the correlation
function.  It is commonly assumed that this measurement is drawn from a
multi-variate Gaussian distribution with a given mean $\left<{\bf D}\right>$
and covariance matrix, $\mathbfss{C}$.  Following a Bayesian framework, the
measurement ${\bf D}$ can be used to constrain a set of cosmological parameters
${\bm \theta}$ by means of an unbiased model ${\bf T}({\bm \theta})=\left<{\bf
D}\right>({\bm \theta})$.  In such way, the probability that the data vector
$\mathbf{D}$ corresponds to a realization of the model $\mathbf{T}({\bm
\theta})$ is given by the likelihood function
\begin{equation}
\mathcal{L}(\mathbf{D}|{\bm \theta},\mathbf{\Psi})
\propto |\mathbf{\Psi}|^{1/2}\mathrm{exp}
\left[
	-\frac{1}{2}\chi^2(\mathbf{D},{\bm \theta},\mathbf{\Psi})
\right],
\label{likl}
\end{equation}
where $\chi^2$ is a quadratic form
\begin{equation}
	\chi^2 = \sum_{ij} \left(D_i - T_i\left({\bm \theta}\right)\right)
	\Psi_{ij} \left(D_j - T_j\left({\bm \theta}\right)\right),
\end{equation}
and $\mathbf{\Psi}$ corresponds to the inverse of the covariance matrix
$\mathbfss{C}$, known as the precision matrix.

The evaluation of the likelihood function requires the knowledge of the
precision matrix, which is commonly derived from a set of $N_{\rm s}$
independent synthetic measurements, ${\bf D}^k$, based on mock catalogues
matching the properties of the real data.  The covariance matrix of the sample
can be inferred using the unbiased estimator
\begin{equation}
	\hat{C}_{ij}= \frac{1}{N_{\rm s}-1} \sum_{k=1}^{N_{\rm s}} 
	(D^k_i-\bar{D}_i)(D_j^k-\bar{D}_j),
\label{eq:cov_est}
\end{equation}
where $\bar{D}_i = \frac{1}{N_{\rm s}} \sum_k D^k_i$ is the mean value of the
measurements at the $i$-th bin, over the set of mock catalogues, which provides
an unbiased estimate of the ensemble average $\left<{\bf D}\right>$.  When
independent realizations are used, the statistics of the uncertainties in the
covariance and precision matrices are governed by the Wishart and
inverse-Whishart distributions \citep{1928Wishart}, respectively.  As the
inverse-Whishart distribution is asymmetric, the inverse of
$\hat{\mathbfss{C}}$ given by equation~(\ref{eq:cov_est}) provides a biased
estimate of the precision matrix.  However, it is possible to correct for this
bias simply by including a prefactor as \citep*{kaufman1967,2007Hartlap}
\begin{equation}
	\hat{\mathbf{\Psi}}=\left(1-\frac{N_{\rm b}+1}{N_{\rm s}-1}\right)
	\hat{\mathbfss{C}}^{-1},
	\label{std_psi}
\end{equation}
where $N_{\rm b}$ corresponds to the number of bins in the measurement ${\bf D}$.

As this approach is based on a finite number of realizations, the estimator in
equation~(\ref{std_psi}) will be affected by noise \citep{2013Taylor}, whose
effect must be propagated into the obtained constraints on the target
parameters.  The accuracy of the obtained confidence levels on the parameters
${\bm \theta}$ and their respective covariances is then ultimately limited by
the uncertainties in the estimated precision matrix $\mathbf{\hat{\Psi}}$
\citep{2014Taylor}. 

\citet{2013Dodelson} performed a detailed analysis of the impact of the
uncertainties in $\hat{\Psi}$.  They showed that, up to second order in the
covariance errors, this additional
uncertainty can be described by a rescaling of the parameter covariances
$\left<\Delta\theta_i\Delta\theta_j\right>$ by a factor
\begin{equation}
f=1+B\left(N_{\rm b}-N_{\rm p}\right),
\label{eq:corr_dodelson}
\end{equation}
where $N_{\rm p}$ corresponds to the number of parameters measured and 
\begin{equation}
B = \frac{(N_{\rm s}-N_{\rm b}-2)}{(N_{\rm s}-N_{\rm b}-1)(N_{\rm s}-N_{\rm b}-4)}.
\label{eq:B}
\end{equation}

However, as shown by \citet{2014Percival}, the correction factor of
equation~(\ref{eq:corr_dodelson}) cannot be directly applied to the errors
derived from maximum likelihood estimates (MLE) based on a given data set.  The
error in the precision matrix introduces a bias in the recovered parameter
uncertainties, which then deviate from those of the ideal case in which the
true covariance matrix is known.  To take this fact into account, the parameter
covariances recovered from the measurements ${\bf D}$ must be rescaled by a
factor
\begin{equation}
g = \frac{1+B(N_{\rm b}-N_{\rm p})}{1 +A+B(N_{\rm p}+1)}\,,
\label{eq:corr_percival}
\end{equation}
where $B$ is given by equation~(\ref{eq:B}) and
\begin{equation}
A = \frac{2}{(N_{\rm s}-N_{\rm b}-1)(N_{\rm s}-N_{\rm b}-4)}. \label{eq:A}
\end{equation}

\citet{2014Taylor} derived general formulae for the full propagation of the noise due to the finite
sampling of the data covariance matrix into the parameter covariance estimated from the likelihood width
and peak scatter estimators, which do not coincide unless the data covariance is exactly known.
Their results are in good agreement with the second-order approximations of 
\citet{2013Dodelson} and \citet{2014Percival} in the regime of $N_{\rm s} \gg N_{\rm b} \gg N_{\rm p}$,
but show deviations from these results for smaller number of simulations. 

In the large $N_{\rm s}$ limit, the correction factor of equation~(\ref{eq:corr_percival})
can be used to obtain constraints that correctly account for the additional
uncertainty due to the noise in the precision matrix estimate of
equation~(\ref{std_psi}). 
However, this additional uncertainty in the parameter covariance
matrix, hinders the constraining power of the data. Given the number of bins
in a measurement ${\bf D}$ and the number of parameters that one wishes to
explore, the correction factor of equation~(\ref{eq:corr_dodelson}) can be used
to estimate the number of synthetic measurements required to reach a given
target accuracy in the derived constraints. If the number of bins in a given
measurement is large, as could be the case for anisotropic or tomographic
clustering measurements, the required number of mock realizations to keep the
additional uncertainty under control might become infeasible.  The aim of our
analysis is to test a new technique to reduce the impact of the uncertainties
in ${\bf \hat{\Psi}}$ on the final parameter constraints, which could help to
significantly relieve these requirements. 

\section{Covariance Tapering for likelihood-based estimation} 
\label{taper}

\begin{figure}
\includegraphics[width=0.45\textwidth]{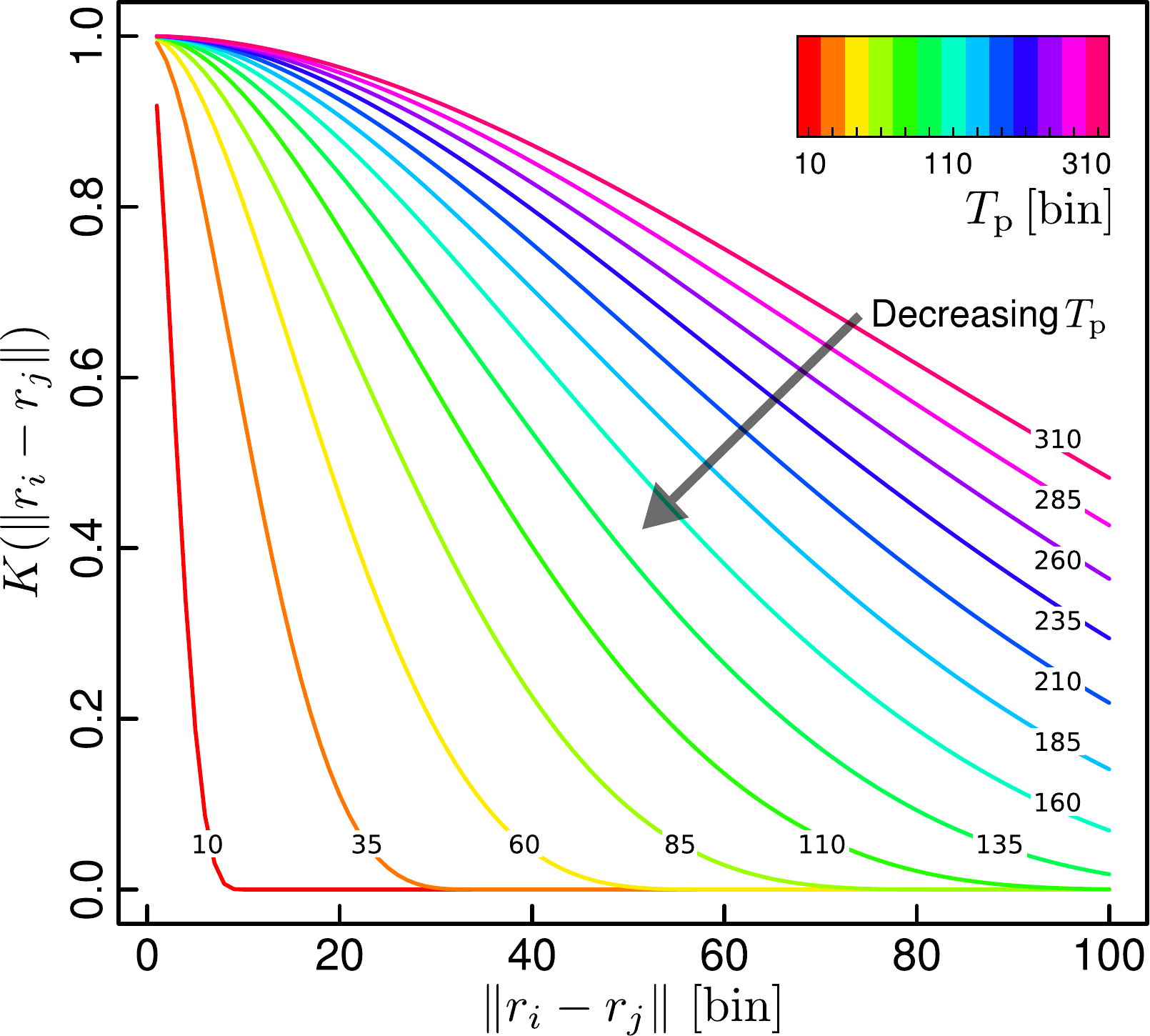}
\caption{
 The behaviour of the family of tapering functions used in this work (the
 Wendland 2.0 function class) corresponding to different values of the
 tapering parameter $T_{\rm p}$. 
 The abscissas represent the distance between two measurement locations
 (denoted by $r_i$) in data set space, which are shown in units of data
 ordinals (i.e bin units, $r_i\equiv i$).
 Larger values of $T_{\rm p}$ result in functions with a larger support interval. 
}
\label{fig_0}
\end{figure}

In this section we describe the covariance tapering technique developed by
\citet{kaufman_covariance_2008} to improve the efficiency on the computation of MLE. This method was
first applied to clustering measurements in \citet{2013Paz}. The original idea behind CT
is the fact that, in many applications, the correlation between data pairs far apart is negligible
and little information is lost by treating these points as being independent.  In this case, setting
their corresponding elements in the covariance matrix to be exactly zero makes it possible to take
advantage of fast numerical methods for dealing with sparse matrices, leading to a significant speed
up of the evaluation of the likelihood function.

However, in this work we focus on a different use of CT.  The off-diagonal elements of the
covariance matrix of a general measurement might exhibit a wide range of values and uncertainties.
Even when these elements are non-negligible, their relevance to obtain an accurate description of
the likelihood function must be assessed in terms of their associated errors. In this way, by
down-weighting the contribution of these points (which typically have a low signal to noise ratio)
to the estimated precision matrix it is possible to avoid the propagation of errors into the final
cosmological constraints. 

\citet{kaufman_covariance_2008} define a tapered covariance matrix, $\mathbfss{C}^{\rm t}$, in terms
of the estimate $\hat{\mathbfss{C}}$ of equation~(\ref{eq:cov_est}) and a properly defined tapering
matrix $\mathbfss{T}$ as
\begin{equation}
	\mathbfss{C}^{\rm t}=\hat{\mathbfss{C}}\circ \mathbfss{T},
\end{equation}
where $\circ$ indicates the Hadamard product (i.e. the entry-wise product). The Schur product
theorem guaranties that if $\mathbf{\hat C}$ and $\mathbfss{T}$  are positive definite matrices, then
so is $\mathbfss{C}^{\rm t}$.  The tapering matrix $\mathbfss{T}$ is defined as an isotropic
covariance matrix by means of a taper function $K$ as
\begin{equation}
T_{ij}= K(\Vert r_i-r_j \Vert),
\end{equation}
where $r_i$ is the $i$-th measurement location on the data space (e.g. the bin separation for
correlation function measurements).  The function $K$ could in principle be any positive
compact-support function intended to nullify the covariance matrix entries far away from the
diagonal. However only certain types of functions (the Mat\'ern class) ensure the asymptotic
convergence of the method to the desired maximum likelihood estimate \citep[see theorems 1 and 3
in][]{kaufman_covariance_2008}. Following \citet{kaufman_covariance_2008} we use the monoparametric
family of functions defined by \citet{wendland_a,wendland_b}
\begin{eqnarray}
K(x)=\left\{\begin{array}{ll}
		\left(1-\frac{x }{T_{\rm p}}\right)^4 \left(4\frac{x}{T_{\rm p}}+1\right)
		&\mathrm{if}\;  x  < T_{\rm p}\\
		0 &\mathrm{if}\;  x  \ge T_{\rm p}
	\end{array}\right.
\end{eqnarray}
Fig. \ref{fig_0} shows the behaviour
of these functions for different values of the tapering parameter, $T_{\rm p}$, which defines the
size of the function support \citep[i.e.  the interval where $K$ takes non-zero values, for more
details see][]{kaufman_covariance_2008}.

As shown by \citet{kaufman_covariance_2008}, the estimation of the precision matrix simply as the
inverse of the tapered covariance $\mathbfss{C}^{\rm t}$ can introduce systematic biases on the
obtained parameter constraints.  However, by applying a second Hadamard product to the inverse of
$\mathbfss{C}^{\rm t}$ and estimating the precision matrix as
\begin{equation}
	\mathbf{\Psi^\mathrm{t}}=\left(1-\frac{N_{\rm b}+1}{N_{\rm s}-1}\right)
	\left(\hat{\mathbfss{C}}\circ\mathbfss{T}\right)^{-1}
	\circ \mathbfss{T}
	\label{psi2taper}
\end{equation}
it is possible to obtain an unbiased and robust estimate of the precision matrix over a large range
of $T_{\rm p}$ values.
  
The likelihood is then estimated by replacing $\mathbf{\Psi}$ in equation \ref{likl}, by the
two-tapered precision matrix estimator,$\mathbf{\Psi^\mathrm{t}}$.  In the following sections we will show
that the CT method is not only useful to approximate the MLE in a more computationally efficient
way, as shown in \citet{kaufman_covariance_2008}. As we will see, CT is also an appropriate
technique to significantly reduce the impact of the noise in the precision matrix estimated from a
set of mock measurements, increasing in this way the precision of the obtained likelihood confidence
regions.

\begin{figure} 
 \includegraphics[width=0.47\textwidth]{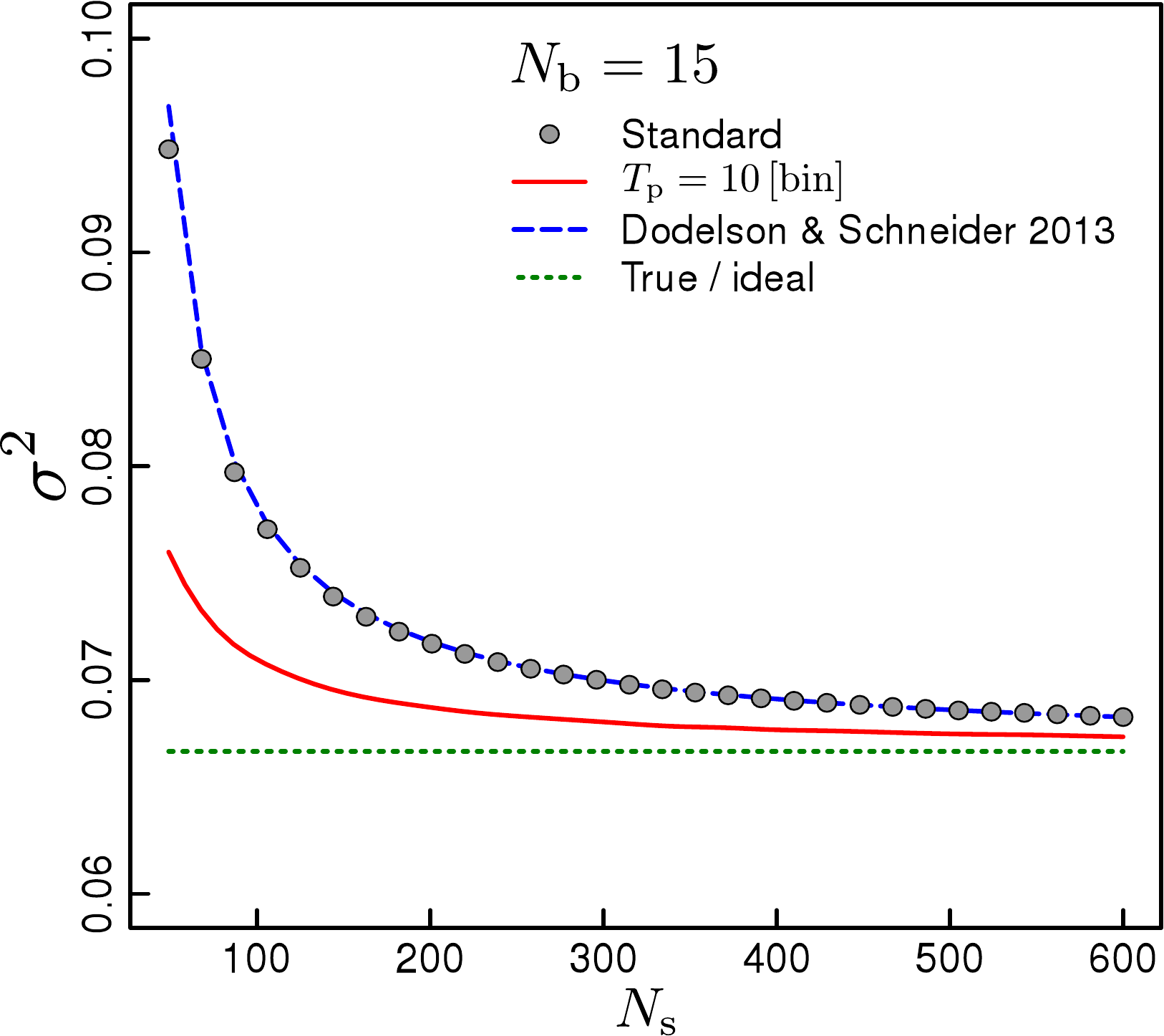} 
 \caption{ 
	Variance of $10^5$ MLE of the mean of sets of $N_{\rm b}=15$ uncorrelated normal
	random variables, as a function of the number of independent data realizations used
	on estimation of the precision matrix. Open symbols correspond to the variance
	obtained when the standard method is used (i.e.  the precision matrix is
	approximated by $\mathbf{\hat{\Psi}}$). The dashed line shows the results obtained
	when applying the CT technique (i.e. the precision matrix is approximated by
	$\mathbf{\Psi^\mathrm{t}}$) with a tapering parameter of $10$ bins. The solid line
	corresponds to the analytic formulae given in \citet{2013Dodelson}. The dotted line
	correspond to the expected variance for the mean of uncorrelated Gaussian random
	variables with first and second moment equal to zero and one, respectively.
 } \label{fig2} 
\end{figure}

\section{Covariance Tapering in practice}
\label{ct_practice}

\subsection{Testing CT on independent normal-distributed measurements}
\label{uncorr}

In this section we compare the performance of the CT approach with the standard MLE method when
applied to the case of $N_{\rm b}$ independent normal-distributed random variables, in a similar
analysis to those performed by \citet{2013Dodelson,2013Taylor} and \citet{2014Percival}.  This
simple test is able to illustrate how the CT technique can be used to minimize the effects of
covariance errors on the estimation of likelihood confidence intervals. 

Our data set for this test consists of $N_{\rm b}$ Gaussian numbers with null mean and standard
deviation equal to unity, in which case the covariance matrix is given by the $N_{\rm b}\times
N_{\rm b}$ identity matrix.  However, we perform a MLE of the sample mean, $\mu$, without assuming
any data independence.  We first compute $\hat{\mathbfss{C}}$ from a set of $N_{\rm s}$ independent
Monte-Carlo realizations of $N_{\rm b}$ independent Gaussian numbers $D^s_i$.  This estimate can be
used to obtain $\mathbf{\hat\Psi}$ and $\mathbf{\Psi^t}$ using eqs.(\ref{eq:cov_est}) and
(\ref{psi2taper}).  Both of these estimates will include ``apparent'' correlations between different
bins, due to the noise in the off-diagonal elements.  We generate an additional independent set of
$N_b$ Gaussian numbers to be used as the data set for the target parameter estimation.  The
estimation of the sample mean is achieved by maximizing the likelihood function in
equation~(\ref{likl}). In this case the model is quite simple, a constant function ${T_i(\mu)}=\mu$.
This procedure is repeated $10^5$ times, obtaining an estimation for the target parameter on each
time.  By using the set of all the estimated values of $\mu$ we are able to compute the standard
error $\sigma$ achieved by the MLE method. The use of an independent data set on parameter
estimation, employing the first $N_{\rm s}$ samples for the estimation of $\hat{\mathbfss{C}}$,
gives an unbiased set of target parameter estimations \citep{2014Percival}.  

The results of this test are shown in Fig. \ref{fig2}, where the open points
correspond to the parameter variance obtained when the standard technique is
used, i.e.  the precision matrix is approximated by $\mathbf{\hat{\Psi}}$, for
the case of $N_{\rm b}=15$.  The comparison of these results with the dotted
line, which corresponds to the true expected variance of the mean of a Gaussian
random variable, illustrates the effect of the noise in the covariance matrix.
As can bee seen, the variance decreases as $N_{\rm s}$ increases, which is
expected due to the corresponding improvement on the $\hat{\mathbfss{C}}$
estimations. This behaviour is well described by the formulae given in
\citet{2013Dodelson}, as can bee seen by looking at the dashed line in Fig.
\ref{fig2}. The same behaviour has been seen over a wide range of $N_{\rm b}$. 
The agreement found here is consistent with the
results of \citet{2014Taylor}, given that the cases considered here correspond to the regime of
large $N_{\rm s}$ compared to the number of parameters and data bins used. 
The solid line
corresponds to the variance obtained when the precision matrix is
estimated using $\mathbf{\Psi^\mathrm{t}}$ of equation
(\ref{psi2taper}) with a tapering scale of 10 bins.  The application of
the CT method significantly reduces the impact of the noise in the
variance of the target parameter, leading to results that are much
closer to those of the ideal case.


\begin{figure*} \includegraphics[width=0.9\textwidth]{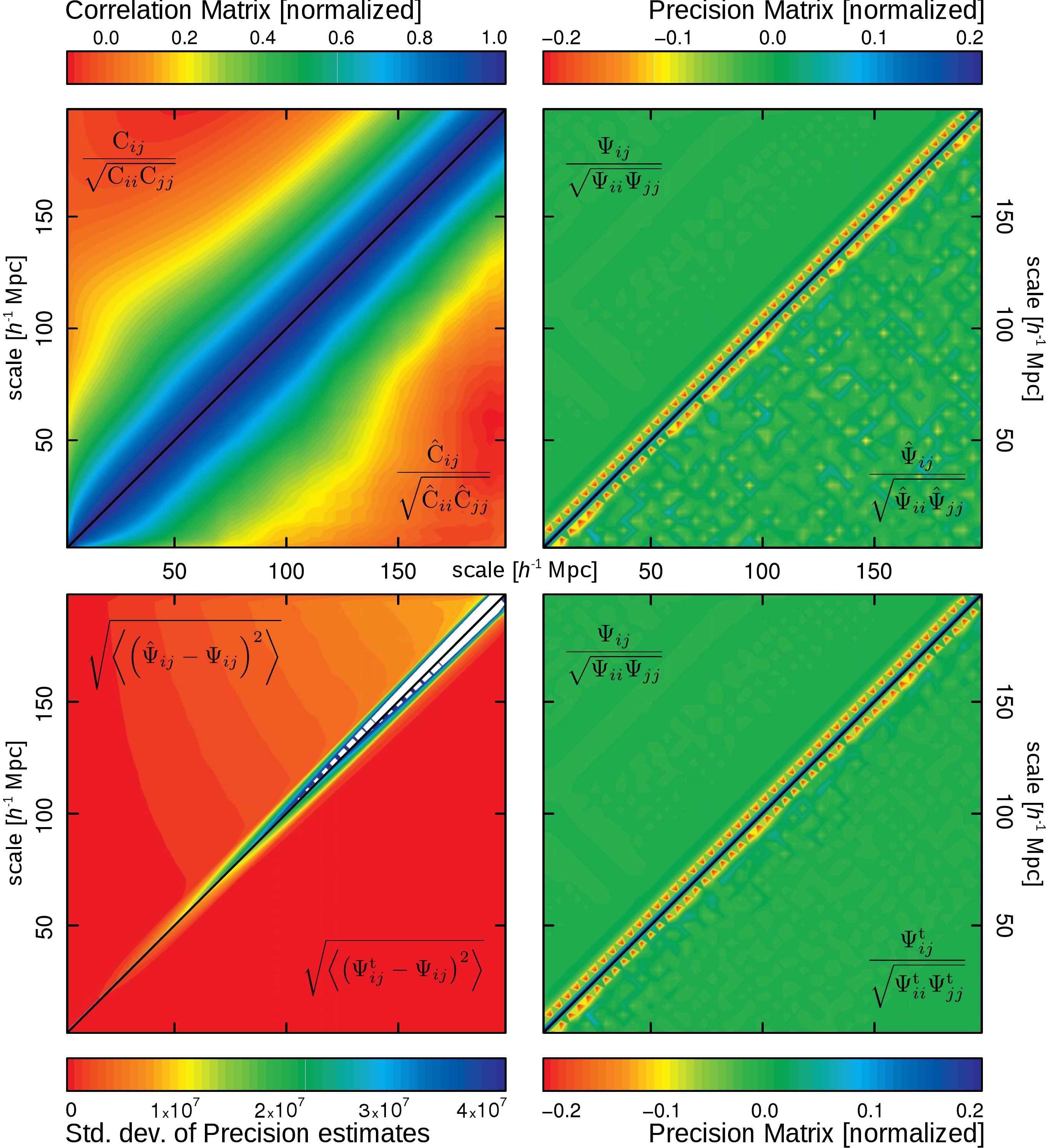} 
\caption{
	{\it Top left panel:} Comparison between the model for the covariance matrix used in this
	work (upper triangular part) and the standard estimation using $300$ realizations. {\it Top
	right panel:} The true precision matrix $\mathbf{\Psi}$ (the matrix inverse of the model,
	upper triangular part) compared to the usual estimator $\mathbf{\hat{\Psi}}$. {\it Bottom
	right panel:} As in the above panel, but comparing this time $\mathbf{\Psi}$ with the CT
	estimator $\mathbf{\Psi^\mathrm{t}}$ using $T_{\rm p}=230\,\mpc$.  {\it Bottom left panel:} Standard deviations for
	$10^4$ precision matrix estimates, with (lower triangular part) and without CT (upper
	triangular part), around the true matrix $\mathbf{\Psi}$.
} \label{fig_3}
\end{figure*}

\subsection{Testing the CT method for realistic covariances} \label{cov_est}

In the previous section we tested the results obtained by applying CT in an ideal case in which the
true covariance matrix is diagonal. In this section we extend this test by considering the case in
which the different elements of the dataset have non-negligible correlations.  To this end, we
generate Monte Carlo realizations of $N_{\rm b}$ Gaussian random variables with zero mean,
correlated following a realistic model of the covariance matrix of the two-point correlation
function \citep{2009Sanchez}. The upper triangular of the top-left panel of Fig. \ref{fig_3} shows
the model for the covariance matrix for the case of $N_{\rm b}=50$, normalized as indicated in the
figure key (that is, the correlation matrix).  For comparison, the lower triangular part of the same
panel shows the estimate $\hat{\mathbfss{C}}$ obtained using $300$ independent Monte Carlo
realizations.  As can be seen, the main features of the model covariance matrix are recovered.
However the presence of noise is clear in the off-diagonal elements of the matrix, corresponding to
covariances of measurements at large separations. The signal-to-noise ratio of the covariance matrix
is smaller for the off-diagonal elements.  The main idea of this work is to control the propagation
of these errors into the precision matrix, restricting in this way their impact on the likelihood
function and the obtained confidence intervals of the target parameters. 

The top right panel of Fig. \ref{fig_3} shows the precision matrices corresponding to the model for
the covariance matrix (upper triangular part) and the one obtained using the standard estimator of
equation (\ref{std_psi}), normalized in analogous manner to the covariance matrices. As can be seen,
the presence of noise in the off-diagonal elements is even more remarkable in the case of the
precision matrix. The increment on the noise is due to the propagation of errors during the
matrix inversion operation.  The bottom-right panel of Fig. \ref{fig_3} shows a comparison of the
precision matrix obtained by applying CT with a tapering parameter $T_{\rm p}=230\,\mpc$
(lower triangular part) and the model precision matrix (upper triangular part, identical to the one
shown in the upper part of the top-right panel). The application of CT leads to a significant
suppression of the noise in the off-diagonal elements of the precision matrix.

The improvement in the accuracy obtained by applying CT can be quantified by
computing the deviations of the different estimators from the true model
precision matrix.  The lower left panel of Fig. \ref{fig_3}, shows the standard
deviations, element by element, of the estimators $\mathbf{\hat{\Psi}}$ and
$\mathbf{\Psi^t}$ for $N_\mathrm{s}=300$, with respect to the ideal precision
matrix. The upper triangular part corresponds to the deviations obtained using
the standard method \citep[which are well approximated by the analytic formulae
given in][]{2013Taylor}, whereas the lower triangular part shows the results
obtained using the CT technique. The standard deviations obtained when CT is applied are
smaller than those achieved by the standard technique, indicating a better performance at
recovering the correct underlying precision matrix. 

\begin{figure}
\includegraphics[width=0.45\textwidth]{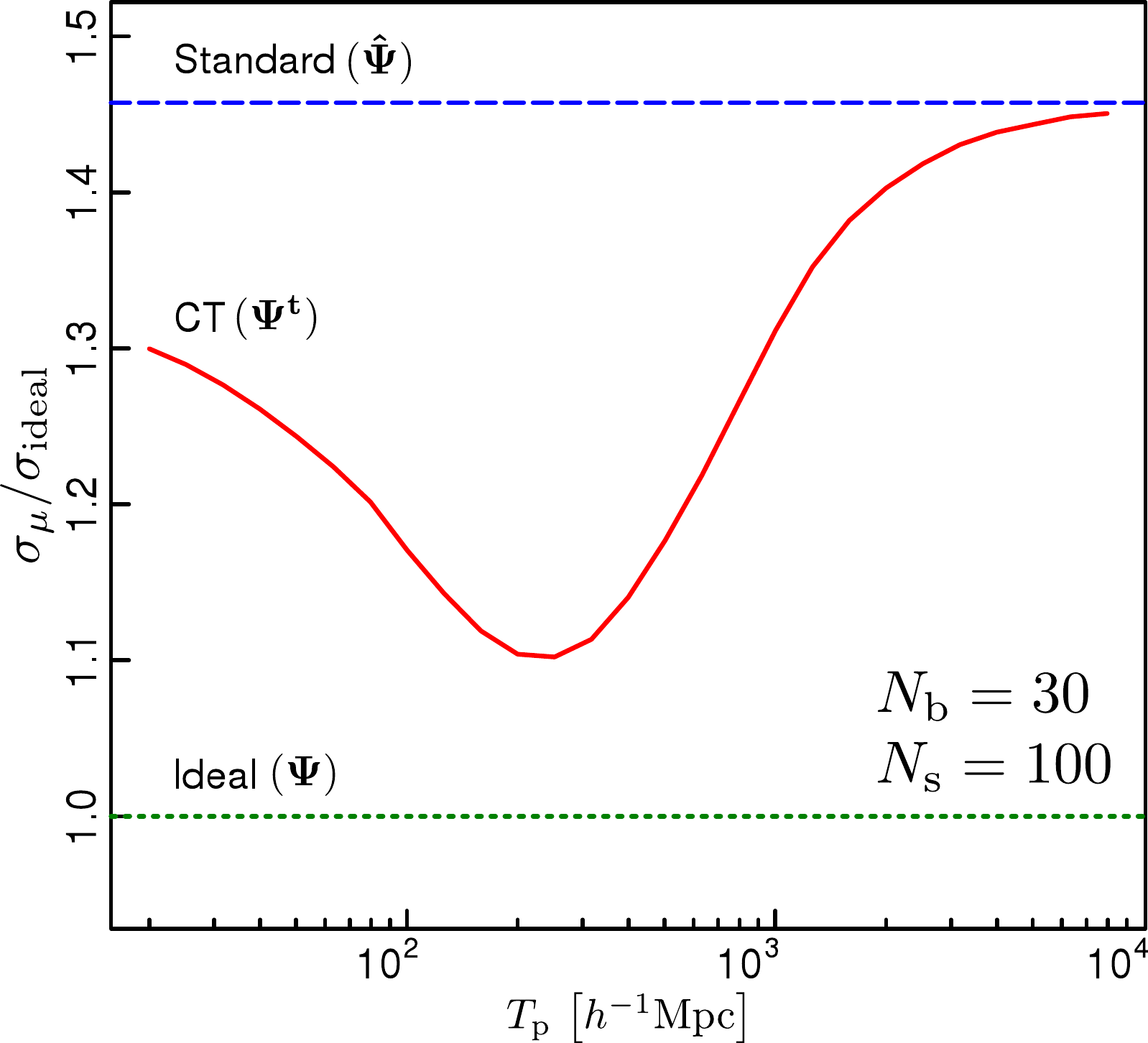}%
\caption{
	 Error of the mean of $N_{\rm b}=30$ random Gaussian numbers with zero mean and covariance
	 given by the model of \citet{2009Sanchez}, normalized to the ideal error
	 $\sigma_\mathrm{ideal}$ predicted by the model. The dashed line corresponds to the error
	 obtained using the standard method, where the $\mathbf{\hat{\Psi}}$ estimator is taken as a
	 proxy of the ideal precision matrix.  The solid line shows the results corresponding to the
	 CT technique, where the precision matrix is estimated by $\mathbf{\Psi^\mathrm{t}}$. The dotted line
	 corresponds to the ideal error.
}
\label{fig_4}
\end{figure}

The success of the covariance tapering method depends on the selection of an adequate tapering scale
$T_{\rm p}$.  In the case analysed in this section, a natural way to characterize the different
approaches is through the recovered error in the mean $\mu$.  The ideal error, which represent the
true constraining power of a given data set, can be easily computed by taking the inverse of the
single element of the fisher matrix. As the derivative of the model with respect to the target
parameter is 1 for each of the $N_{\rm b}$ dimensions, the ideal error of the mean is simply given
by $\sigma_\mathrm{ideal}=1/\sum_{ij} \Psi_{ij}$.  Fig. \ref{fig_4} shows a comparison between the
standard and CT methods for a fixed number of bins and simulations ($N_{\rm b}=30$, $N_{\rm
s}=100$).  For each method we show the ratio $\sigma_\mu/\sigma_\mathrm{ideal}$. The solid line
corresponds to the results obtained by applying CT as a function of the tapering scale $T_{\rm p}$.
The dashed line shows to the error obtained using the standard method, where the precision matrix is
estimated by $\mathbf{\hat{\Psi}}$. The standard deviation obtained in this case corresponds to an
excess of 45\% with respect to that of the ideal case, indicated by the dotted line.  As can be
seen, there is an optimal tapering scale of $T_{\rm p}\simeq230\,\mpc$), for which the error
obtained is only 10\% larger than the ideal value. For large tapering scales, the CT method recovers
the same results as the standard technique, whereas for small $T_{\rm p}$ the errors become larger
again.
This behaviour might be given by the relation between the tapering scale and the structure of the
off-diagonal elements of the covariance matrix. For large values of $T_{\rm p}$, the tapering procedure damps
the contribution of the most off-diagonal elements of the covariance matrix, whose intrinsic values are small
and are dominated by noise. As the tapering scale is reduced, the damping of the noise is more efficient and
the results become more similar to those of the ideal case. However, if the tapering scale is too small,
this procedure might affect entries of the covariance matrix whose intrinsic values are not small.
As these off-diagonal elements are affected, the obtained results deviate again from those of the ideal case.

\begin{figure*}
\includegraphics[width=\textwidth]{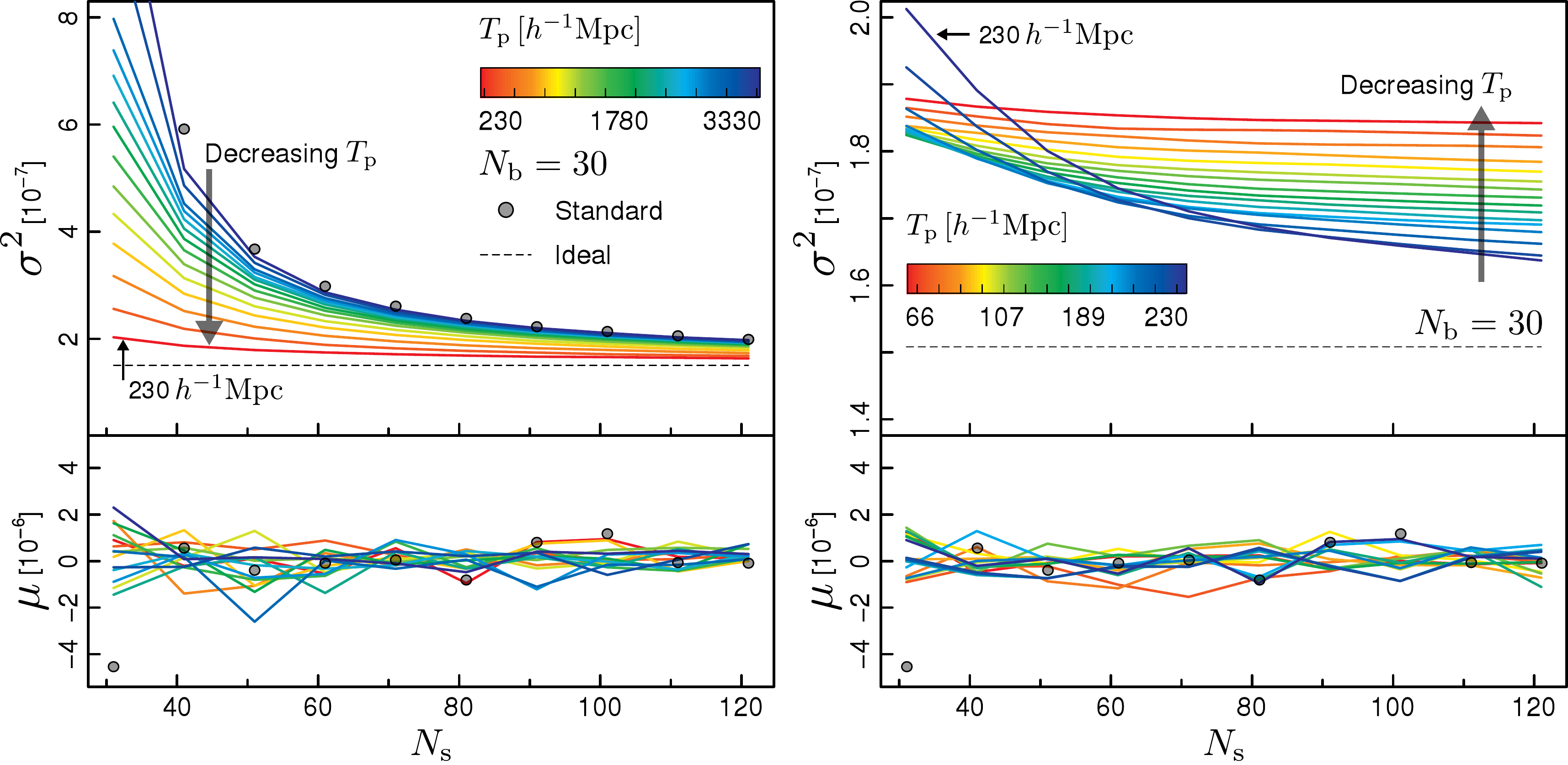}%
\caption{
	Variance of $10^5$ MLE of the mean of $N_b$ Gaussian numbers, following a realistic model of
	the covariance. The results are presented as a function of the number of samples used in the
	covariance estimation. Open symbols correspond to the variance obtained when the precision
	matrix is estimated through the standard method (i.e.  $\mathbf{\hat{\Psi}}\rightarrow
	\mathbf{\Psi}$).  Solid coloured lines show the results obtained by CT using different $T_p$
	parameter values, as indicated in the key.  The dashed line correspond to the expected
	variance for the mean, that is the inverse of the fisher element. Left panel corresponds to
	$T_p$ values between $230$ and $3330\,\mpc$, whereas right panel shows the results of
	$T_p$ ranging from $66$ to $230\,\mpc$.  The bottom subpanels at both sides show the
	recovered $\mu$ values for all methods.
}
\label{fig_5}
\end{figure*}

In Fig.~\ref{fig_5} we show the results of extending this test to a wide range of $N_{\rm s}$
values.  The variance of the MLE of the mean of $10^5$ independent data sets of $N_{\rm b}=30$
correlated Gaussian numbers is shown as a function of the number of realizations employed in the
estimation of $\hat{\mathbfss{C}}$. The points correspond to the variance inferred using the
standard technique, whereas the solid coloured lines indicate the results obtained by applying CT
with different $T_{\rm p}$ values, as indicated in the figure key. The dashed line corresponds to
the expected ideal variance of the mean, $\sigma^2_\mathrm{ideal}$. As can bee seen in the left
panel of this figure, the variance recovered by applying CT with large tapering scales is very
similar to the result of the standard method.  However, as the tapering scale decreases, the
variance also becomes smaller, reaching values close to those of the ideal case for $T_{\rm p}\simeq
230\,\mpc$.  The lower panels in Fig.~\ref{fig_5} show the behaviour of the estimated values of the
target parameter for the standard (points) and CT (solid lines) methods.  In all cases the recovered
values of $\mu$ show no indication of a systematic deviation from the true value $\mu=0$, with only
small fluctuations for different values of $N_{\rm s}$.

The right panel of Fig.~\ref{fig_5} shows the results obtained by applying the CT technique for
values of the tapering scale of less than $230\,\mpc$, which result in larger variances
of the target parameter. However, it is worth noticing that even in this case there is no bias in
the obtained constraints. These results suggest that the optimal tapering scale depends only on the
shape of the underlying covariance matrix, rather than the number of independent data samples used
to compute $\mathbf{\hat{C}}$. 

\section{Application to Baryon Acoustic Oscillation measurement}
\label{app_bao}

In this section we analyse the applicability of the covariance tapering method, in the context of
baryon acoustic oscillations (BAO) measurements.  For this test we use the model of the full shape
of the large-scale two-point correlation function, $\xi(s)$, of \citet{2013Sanchez,2014Sanchez},
which is based on renormalized perturbation theory \citep{2006Crocce}. We generate sets of
correlated Gaussian numbers with mean given by the model of $\xi(s)$ and the same covariance matrix
as in the test of the previous section \citep{2009Sanchez}.  In this way, each Monte Carlo
realization mimics a realistic measurement of the correlation function at the scales relevant for
BAO measurements.

As in previous sections, we generate sets of $N_{\rm s}$ random realizations of the correlation
function, sampled using $N_{\rm b}=30$ bins, varying $N_{\rm s}$ between 40 and 420. We use these
synthetic samples to obtain an estimate of the covariance matrix. These realizations are analogous
to the measurements from mock galaxy catalogues used to compute $\hat{\mathbfss{C}}$ in most
clustering analyses.  We generate an additional synthetic measurement to serve as the objective data
sample of the two-point correlation function, on which we perform fits of the BAO signal following
the methodology of \citet{2014Anderson}. More precisely, we fit the BAO peak position using a
parametrization for $\xi(s)$ given by
\begin{equation}
	\xi_{\mathrm{mod}}(s)=b^2\xi_{\mathrm{temp}}(\alpha s)+a_0 +\frac{a_1}{s}+\frac{a_2}{s^2},
\end{equation}
where $\xi_{\mathrm{temp}}(s)$ is a template given by the model of the correlation function, the
parameter $b$ corresponds to a large-scale bias factor, the shift parameter $\alpha$ is used to
control the position of the BAO peak, and $a_0$, $a_1$ and $a_2$ are additional parameters used to
marginalize over the broad-band signal.  These fits are performed using the standard estimation of
the precision matrix, $\mathbf{\hat{\Psi}}$, and the CT estimate, $\mathbf{\Psi}^{\rm t}$, with
varying tapering scales.  We explore the parameter space ${\bm \theta}= \left( \alpha, b, a_0, a_1,
a_2 \right)$ using the Markov chain Monte Carlo (MCMC) technique. We focus here on the constraints
on $\alpha$, marginalizing over the remaining parameters. 

\begin{figure}
\includegraphics[width=0.47\textwidth]{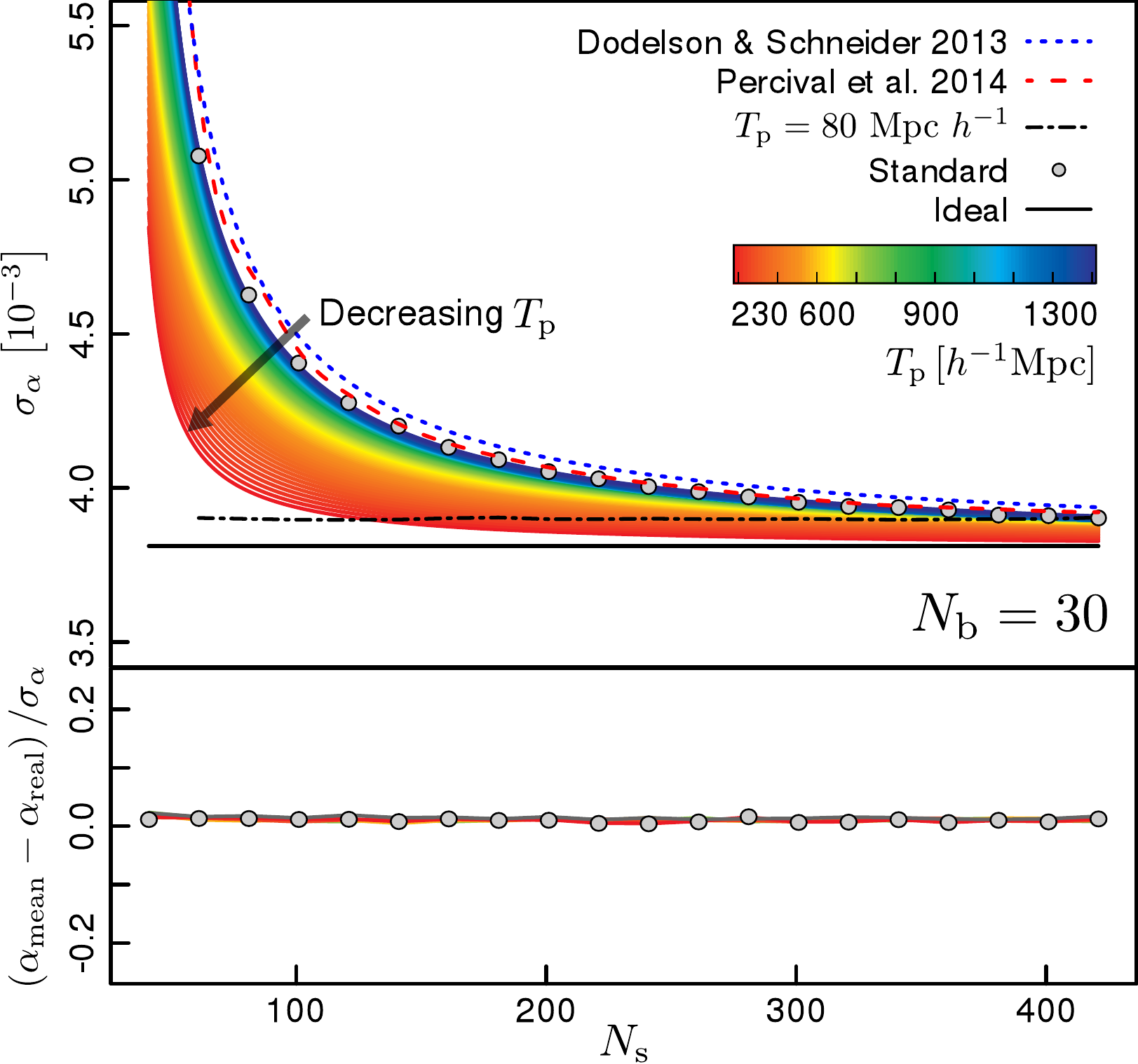}
\caption{
	Standard deviation of the BAO peak scale as a function of the number of simulations used in
	the covariance estimate (upper panel). Open circles indicate the results corresponding to
	the standard method. Long and short dashed lines corresponds to the analytic formulae of
	\citet{2013Dodelson} and \citet{2014Percival}. The black solid line indicates the error
	obtained when the ideal precision matrix is used. Coloured lines (from blue to red) show the
	results obtain by applying CT with different tapering scales.  The black dot-dashed line
	show the CT results when a too small tapering parameter is used ($T_{\rm p}=80\,\mpc$).
}
\label{fig_6}
\end{figure}

The procedure described above is repeated $2\times 10^5$ times to obtain a smooth measurement of the
dispersion of the $\alpha$ values obtained from different realizations. Fig.  \ref{fig_6} shows the
behaviour of this dispersion as a function of the number of mock samples used in the estimation of
the covariance matrix. The points correspond to the results obtained when the precision matrix is
approximated by the estimator $\mathbf{\hat{\Psi}}$ of equation~(\ref{eq:cov_est}).  As expected,
the error in $\alpha$ decreases as $N_{\rm s}$ increases, approaching the ideal error obtained when
the true covariance matrix is used to evaluate the likelihood function, shown by the black solid
line.  The results from our Monte Carlo realizations are well described by the analytic formulae
given by \citet{2013Dodelson}, which is shown by the blue short-dashed line.  The red long-dashed
line corresponds to the result of rescaling the mean variance on $\alpha$ recovered from each MCMC
by the correction factor of equation~(\ref{eq:corr_percival}) which, as shown by
\citet{2014Percival}, correctly accounts for the additional error due to the noise in
$\mathbf{\hat{C}}$.

The thin solid lines in Fig. \ref{fig_6} correspond to the results obtained using the CT method,
colour-coded according to the corresponding tapering scale.  For large tapering scales the results
closely resemble those of the standard technique.  As the tapering parameter decreases, approaching
the optimal scale found in the previous section of $230\,\mpc$, the dispersion of the BAO
scale estimates becomes closer to the ideal error.  These results show that, for a given value of
$N_{\rm s}$, the CT method leads to measurements of the BAO scale that are closer to the true
constraining power of the data than the standard technique. For $N_{\rm s}\simeq 400$ the
uncertainty obtained using the CT method essentially recovers that of the ideal case.

The black dot-dashed line in Fig.~\ref{fig_6} corresponds to the CT results obtained by setting
$T_{\rm p}=80\,\mpc$, illustrating the effect of implementing a too small tapering scale.  As we
found in the previous section, selecting a too small tapering scale leads to an increment of the
errors in the target parameters. However, as shown in the lower panel of Fig.~\ref{fig_6}, even in
this case the CT results show no systematic bias, with negligible differences from the true
underlying parameter $\alpha_\mathrm{true}=1$ for the full range of $N_{\rm s}$ values analysed.

\begin{figure}
\includegraphics[width=0.47\textwidth]{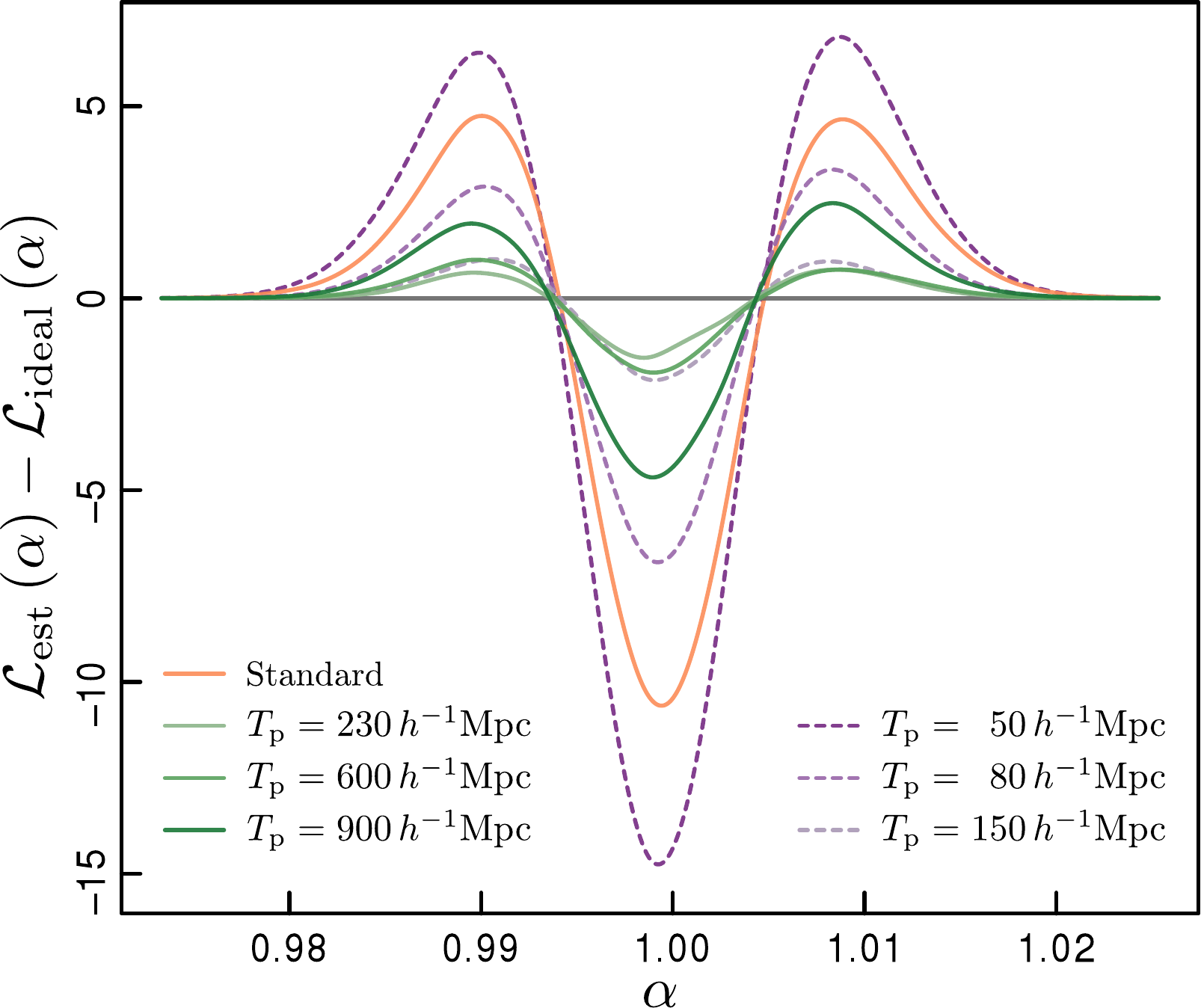}%
\caption{
	Difference between the mean marginalized posterior distribution of the shift parameter
	$\alpha$ obtained by applying CT with different tapering scales (green and purple lines) and
	that of the ideal case (gray line). The orange curve shows the results obtained by the
	standard method. 
}
\label{PDF}
\end{figure}

The improvement of the constraints achieved by the CT technique with respect to the standard method
ultimately relays in a closer approximation of the ideal likelihood surface.  In Fig. \ref{PDF} we
show the difference between the mean marginalized posterior distribution of the shift parameter
$\alpha$ obtained by applying CT with different tapering scales (green and purple lines, as
indicated in the figure key) and that of the ideal case.  The posterior distribution corresponding
to the standard method is shown as a solid orange curve. In general, the CT results provide a closer
approximation of the ideal distribution than the standard method, most notably for a tapering scale
of $230\,\mpc$.  The use of a larger tapering scale leads to larger deviations from the
ideal likelihood function.  However, in all of these cases they are closer to the optimal result
than the standard method.  In contrast, using a too small tapering scales (as in the case of $T_{\rm
p}=50\,\mpc$ shown in the figure) leads to deviations from the underlying likelihood
function that can be even larger than those obtained with the standard technique. This highlights
the importance of performing a careful analysis of the optimal tapering scale for each case, which
will depend on the structure of the covariance matrix. 

\section{Summary and Conclusions}
\label{conclusions}

In this work we have implemented and tested a novel technique for the estimation of the precision
matrix, the covariance tapering method, developed by \citet{kaufman_covariance_2008}.  We have
analysed the performance of the CT method and the standard technique by comparing the obtained
parameter constraints with those found in the ideal case where the true precision matrix is known.
For the latter case, the size of the errors in the estimated parameters is only governed by the
constraining power of the data sample. Therefore, any excess seen in the recovered errors in
comparison with that of the ideal case can be associated with the noise in the precision matrix
estimation.

We first carried out an analysis similar to those performed by \citet{2013Dodelson},
\citet{2013Taylor} and \citet{2014Percival}, using the standard deviation of the maximum likelihood
estimation of the mean of $N_\mathrm{b}$ independent normal random variables as a test case.  Our
results are in agreement with previous works, indicating that the noise in the covariance matrix
estimation has a significant impact on the uncertainty obtained on parameter constraints. We found
that the size of the synthetic data set used to estimate the precision matrix is crucial to control
the impact of the covariance uncertainties on the final parameter constraints. Our results also show
that CT helps to reduce the error in the precision matrix estimation, leading to uncertainties in
the target parameters that are closer to the ideal case than those obtained with the standard
method.

The efficiency of the CT technique for a data set composed of independent normal variables,  where
the true covariance matrix is diagonal, can be expected.  In order to check the validity of the CT
method in a more realistic situation, we also studied the case of the MLE of the mean of Gaussian
numbers with non-zero correlations given by a realistic model of the covariance matrix of the
two-point correlation function \citep{2009Sanchez}.  Our results show that also in this case the
covariance tapering method leads to smaller errors than the standard technique, without introducing
any systematic bias in the estimated parameters.  In this case we found an optimal tapering scale,
defined as the value of the tapering parameter for which the obtained standard deviation is closer
to its ideal value.  For example, in the case of $30$ normal distributed values and $100$ synthetic
samples on the covariance estimation, the CT method with the optimal tapering scale, gives a
standard deviation only 10\% larger than that of the ideal case, whereas the standard technique
results in errors 45\% larger. We showed that the optimal tapering parameter depends only on the
structure of the underlying covariance matrix and is insensitive to the bin size or the number of
synthetic samples used in the estimation of the precision matrix.  Smaller tapering parameters than
the optimal value result in an increase of the standard deviation, although no bias is introduced. 

Finally, we performed an analysis of the CT technique on a more realistic context by testing its
applicability to isotropic BAO measurements.  We used an accurate model of the large-scale
two-point correlation function \citep{2013Sanchez} and its full covariance matrix
\citep{2009Sanchez} to generate Monte Carlo realizations of this quantity. We used these synthetic
measurements to perform fits of the BAO signal following the methodology of \citet{2014Anderson}.
We found that the CT method significantly reduces the impact of the noise in the precision matrix on
the obtained errors in the BAO peak position without introducing any systematic bias.  As in the
previous case, the optimal tapering parameter only depends on the shape of the true covariance
matrix, with a preferred $T_\mathrm{p}$ scale similar to that of the previous test (i.e. the case of
zero-mean correlated Gaussian numbers). 

Covariance tapering can help to reduce the required number of mock catalogs for the analysis of
current and future galaxy surveys. This can be clearly illustrated by extending the analysis of
section \ref{app_bao} to the case of $N_\mathrm{s}=600$, corresponding to the number of mock
catalogues used in the analysis of the SDSS-DR9 BOSS clustering measurements of
\citet{2012Anderson}.  In this case, the uncertainty in the BAO shift parameter obtained by applying
the CT technique is equivalent to that derived with the standard method using $N_\mathrm{s}=2300$
instead.  

As we highlighted before, the performance of the CT technique ultimately depends on the structure of
the underlying covariance matrix. In this work we assumed a Gaussian model for the covariance matrix
of the correlation function. Although this model gives an excellent description of the results of
numerical simulations, more accurate models must include also the contribution from modes larger
than the survey size \citep{2012dePutter} or non-Gaussian terms \citep{1999Scoccimarro} that would
affect the off-diagonal elements of the covariance matrix.  We leave the study of the performance of
the CT under the presence of these contributions for future work, as well as the extension of the
present analysis to alternative data sets such as the power spectrum or anisotropic clustering
measurements in general.

The covariance tapering technique can be extremely useful for the analysis of future surveys such as
Euclid and DESI. The small statistical uncertainties associated with these data sets will provide
strong tests for the standard $\Lambda$CDM cosmological model. However the large number of mock
catalogues that are required by the standard technique to maintain the accuracy level of the
cosmological constraints might be infeasible.  The application of the covariance tapering technique
can significantly reduce the number of mock catalogues required for the analysis these surveys,
allowing them to reach their full constraining power.

\section*{Acknowledgments}

The authors are thankful to Andr\'es Nicol\'as Ruiz and Salvador Salazar-Albornoz for useful
discussions and suggestions about this manuscript.  DJP acknowledges the support from Consejo
Nacional de Investigaciones Cient\'{\i}ficas y T\'ecnicas de la Rep\'ublica Argentina (CONICET,
project PIP 11220100100350)  and the Secretar\'{\i}a de Ciencia y T\'ecnica de la Universidad
Nacional de C\'ordoba (SeCyT, project number 30820110100364).  DJP also acknowledges the hospitality
of the Max-Planck-Institut f\"ur extraterrestrische Physik were part of this work was carried out.
AGS acknowledges the support from the Trans-regional Collaborative Research Centre TR33 `The Dark
Universe' of the German Research Foundation (DFG).


\end{document}